\documentstyle[twoside,fleqn,hyperon99_paper]{article}
\input epsf

\newcommand{\CP}{$CP$}

\newcommand{\AmS}{{\protect\the\textfont2
  A\kern-.1667em\lower.5ex\hbox{M}\kern-.125emS}}

\title{\begin{flushright}
{\normalsize
IIT-HEP-99/2\\
October 1999\\
\vspace{0.1in}}
\end{flushright}
Should ${\bar p}p\to{\bar\Lambda}\Lambda$ be revived?}

\author{Daniel M. Kaplan \address{Physics Division, Illinois Institute of Technology,\\
Chicago, IL 60616, USA
}%
        \thanks{E-mail: kaplan@fnal.gov}\thanks{To appear in {\sl Proc.\ Hyperon
	99}, Fermilab, 27--29 Sept.\ 1999.}
}
\begin{document}

\begin{abstract}
The continued interest of {\em CP} violation in hyperon decay (as well as
many other physics topics that could be addressed by such a facility) suggests that a dedicated $\bar p$ storage ring at Fermilab
ought to be reconsidered. With recent and anticipated technical progress, sensitivity
many orders of magnitude beyond that achieved in LEAR may be possible, including
$10^{-5}$ sensitivity for the $\Lambda/{\bar\Lambda}$ {\em CP}-asymmetry parameter $A_\Lambda$.
\end{abstract}

\maketitle
\setcounter{footnote}{0}

From 1984 until LEAR was shut down in 1996, the reaction ${\bar
p}p\to{\bar\Lambda}\Lambda$ was extensively studied by the  PS185
experiment~\cite{PS185}. This technique was proposed by Donoghue, Holstein, 
and Valencia~\cite{Donoghueetal}
in 1986 (and during 1990--1992 further elaborated  by the CERN CP-Hyperon Study
Group~\cite{CERN-study}) as a possible avenue to detection of {\em CP}
violation in hyperon decay. The PS185 group has since published~\cite{Barnes}
the world's most sensitive limit to date on the $\Lambda/{\bar\Lambda}$ {\em CP}
asymmetry~\cite{Donoghue-Pakvasa} $A_\Lambda\equiv (\alpha+{\bar \alpha})/(\alpha-{\bar
\alpha})=0.013\pm0.022$, based on $\approx10^5$ events, and analysis of the
large PS185 data sample continues.\footnote{Here $\alpha$ is the up-down
asymmetry parameter for $\Lambda\to p\pi^-$ decay and ${\bar\alpha}$ is that for
the charge-conjugate decay.}

The CERN hyperon-{\em CP}-violation study also
stimulated  the 1992 Fermilab Proposal 859 by Hsueh and
Rapidis~\cite{P859}. By this time PS185 had demonstrated ${\cal O}(10^{-2}$)
sensitivity, and the goal of P859 was sensitivity of $1\times10^{-4}$, where
model calculations~\cite{Donoghue-He-Pakvasa} predicted a
possibly-detectable effect. They proposed a modification of the PS185 approach,
with a new dedicated ${\bar p}$ storage ring to be built at Fermilab for the
purpose. The proposal was turned down, with the comment that it would take
$10^{-5}$ sensitivity to justify building a new
storage ring~\cite{Peoples-letter}. A 1993 proposal for a fixed-target
experiment at $1\times10^{-4}$ sensitivity was eventually approved, leading
to the HyperCP experiment now running in the Meson Center beamline~\cite{E871}.

While the large value of $\epsilon^\prime/\epsilon=
(21.2\pm4.6)\times10^{-4}$~\cite{He,eps'}  
suggests the possibility that $A_\Lambda$ might be
similarly large~\cite{He}, whether HyperCP observes a
few$\,\times\,10^{-4}$ to 10$^{-3}$ effect or not, it is of interest whether
sensitivity at the $10^{-5}$ level is feasible.\footnote{Since HyperCP measures the sum
of the $\Xi$ and $\Lambda$ asymmetries, a direct measurement of $A_\Lambda$ 
will be important even if HyperCP observes a large value for the sum
$A_\Xi+A_\Lambda$.} This would require
$\sim10^{11}$ events, 100 times as many as in HyperCP --- probably not
feasible in the fixed-target approach. Thus we should explore whether ${\bar
p}p\to{\bar\Lambda}\Lambda$ could be pushed to 10$^{-5}$.

The P859 sensitivity estimate was based on 3 months of running at an average
luminosity of $1.6\times10^{32}\,$cm$^{-2}$s$^{-1}$, 
to be achieved with 260\,mA of
1.64\,GeV/$c$ antiprotons, in a ring about 1/3 the size of the
Accumulator, incident on a hydrogen-gas-jet  target of
$1\times10^{14}\,$atoms/cm$^2$. This luminosity requires antiproton
production at a minimum rate of $6\times10^{10}$/hour. 

While these numbers were
ambitious for 1992, they have since been surpassed by the $\bar p$
source and the E835 gas-jet
target. A factor 100 in event sample is nevertheless a
tall order. It might be achievable with (for example) luminosity
$1\times10^{33}\,$cm$^{-2}$s$^{-1}$ over four years of running, with 500\,mA
of antiprotons and target density of $3\times10^{14}\,$atoms/cm$^2$. While this
target density has been achieved by the E835 collaboration~\cite{Pordes-private},
the
required antiproton production rate of $3.6\times10^{11}$/hour is a factor
$\approx$2 extrapolation beyond current plans for Tevatron Run
II~\cite{Harms-private}. This is
not unreasonable, especially given plans for a substantial proton-source upgrade
at Fermilab~\cite{Chou}.

Of course, the establishment or refutation of feasibility at $10^{-5}$
sensitivity will require a great deal more work --- 
for example, on storage-ring optics and cooling, triggering and data 
acquisition,
and especially a detailed
study of systematic uncertainties. The importance of this physics suggests that
such an effort is worthwhile. More generally, the availability of a $\bar p$
source orders of magnitude beyond LEAR in intensity should make
possible a wide range of interesting physics. Given the relatively modest cost
of such a project, it should be seriously considered as an add-on to the 
Fermilab program in the coming decade.

\end{document}